\documentclass[12pt]{JHEP}
\usepackage{amsmath,amssymb}
\usepackage{yfonts}

\newcommand{\be}{\begin{equation}}
\newcommand{\ee}{\end{equation}}
\newcommand{\beqa}{\begin{eqnarray}}
\newcommand{\eeqa}{\end{eqnarray}}

\newcommand{\PP}{ {\cal P} } 

\def\bemat{\left( \begin{array}}
\def\enmat{\end{array} \right)}

\newcommand{\qn}{\textswab{q}}
\newcommand{\wn}{\textswab{w}}

\def\med{\frac{1}{2}}

\def\Im{\mathop{\rm Im}}

\newcommand{\HHH}{{\cal H}}

\def\Nfour{{\cal N}\!=\!4}

\relax

\title{The shear viscosity of the non-commutative plasma }

\author{ Karl Landsteiner$^{a}$ and Javier Mas$^{b}$
\\
$^{a}$ Instituto de F\'\i sica Te\'orica  C-XVI, 
Universidad Aut\'onoma de Madrid\\
28049 Madrid, Spain \\
email: {\texttt{karl.landsteiner@uam.es}}

$^{b}$
 Departamento de F\'\i sica de Part\'\i culas,
Universidad de Santiago de Compostela \\ E-15782 Santiago
de Compostela, Spain\\
email: \texttt{javmas@usc.es}
}

\abstract{We compute the shear viscosity of the non-commutative N=4 
super Yang-Mills quantum field theory at strong coupling using the 
dual supergravity background. 
Special interest derives from the fact that the background presents 
an intrinsic anisotropy in space through the distinction of commutative 
and non-commutative directions. Despite this anisotropy the analysis 
exhibits the ubiquitous  result $\eta/s = 1/4\pi$ for two different 
shear channels. In order to derive this result, we show that the 
boundary energy momentum tensor must couple to the open string metric. 
As a byproduct we compute the renormalised holographic energy momentum 
tensor and show that it coincides with one in the commutative theory.
}
\preprint{IFT-UAM/CSIC-07-31 }


\begin{document}

\section{Introduction}

One of the most interesting developments and applications of string
theory in the recent years has been the study of the properties of
strongly coupled gauge theories at high temperatures via the AdS/CFT
duality \cite{Maldacena:1997re}. At sufficiently high temperatures
gauge theories go over into a plasma phase. According do the AdS/CFT
dictionary the plasma phase is represented in the holographic dual as
a gravitational background containing a black hole
\cite{thermal}.

Earlier applications of asymptotically AdS black holes to the study of
strongly coupled gauge theories have focused on the thermodynamics of
the system, typically studying Hawking temperature, entropy, free
energy and phase transitions, which can be studied using the Euclidean section
of the black hole metrics.

The recent interest has however derived from the study of
non-equilibrium dynamics and here it is essential to use the real time
Lorentzian background \cite{Son:2007vk}. One of the main motivations
is the experimental progress in the study of the Quark-Gluon plasma at
the Relativistic Heavy Ion Collider (RHIC) in Brookhaven. It turned
out that the state of matter created at the heavy ion collisions at
RHIC is most likely to be understood as a strongly coupled Quark Gluon
plasma (for a review see \cite{Nagle:2006cj}). This poses great
problems for the theory since usual perturbative field theory techniques
are not applicable and lattice simulations are mostly confined to the
equilibrium regime. Therefore the AdS/CFT duality has emerged as a
useful tool for studying the properties of the plasma phase on
non-abelian gauge theories at strong coupling.

One of the most impressive results in this line of investigation has
been the calculation of the shear viscosity in holographic gauge
theories \cite{Policastro:2001yc}. There are two extremely interesting
aspects of these kind of calculations. One is that the actual
numerical value turns out to be consistent with experiments ( see e.g.
\cite{Muller:2006ee}). The other, more theoretical one, is the fact
that all calculations that have been done so far in different
holographic theories have always produced the same
result for the ration of the shear viscosity over the entropy
\begin{equation}\label{eq:eta_universal}
  \frac{\eta}{s} = \frac{1}{4\pi}\,.
\end{equation}
 This gave rise to the
conjecture that the value is universal and represents a lower bound
for all physical systems \cite{Kovtun:2004de}\footnote{See however
  recent attempts to construct ``gedanken" counterexamples in
  \cite{Cohen:2007qr}.}.  This universality was first suggested
through a case by case investigation, while attempts at a general
proof have always faced some restrictions over the class of metrics
involved \cite{universality}.

As system under consideration we have chosen the holographic dual to
the non-commutative $\Nfour$ gauge theory
\cite{Hashimoto:1999ut,Maldacena:1999mh}. We will see that
understanding holography in this theory is by itself an interesting
endeavour. In view of the calculations of the shear viscosity
additional interest derives from the fact that the non-commutative
theory is anisotropic in space. We will consider the theory with two
space coordinates $(y,z)$ being non-commutative and leave $(t,x)$
usual commutative spacetime coordinates. Now we remember that the
viscosity is actually a fourth rank tensor relating the gradients in
the local fluid velocity to the dissipative part of the stress tensor
\begin{equation}
  T^D_{ij} = \eta_{ij,kl} \frac{\partial u_k}{\partial x_l}\,.
\end{equation}
Assuming the usual $SO(3)$ rotational invariance leaves only two
independent tensor structures, corresponding to shear viscosity $\eta$
and bulk viscosity $\zeta$. In the anisotropic, non-commutative theory
we have however only an $SO(2)$ symmetry and this gives rise to a much
richer tensor structure. We will be interested only in the shear part,
i.e. the part that determines the diffusion of momentum into
transverse directions. More precisely we have to distinguish between
three possible shear viscosity coefficients corresponding to either
momentum in the commutative direction diffusing into a non-commutative
direction, momentum in a non-commutative direction diffusing into a
commutative direction and momentum in a non-commutative direction
diffusing into a non-commutative direction\footnote{Anisotropic shear
  viscosities are well-known in the theory of liquid crystals where
  there is a director field indicating a preferred axes of alignment.
  The three different shear viscosities that parameterise the
  different momentum diffusion processes relative to the director
  field are called Miesowicz coefficients \cite{deGennes}.}.

On a technical level the shear viscosities can be calculated in
various different ways, either by searching for a diffusion pole in
the retarded two point function of the momentum density operators or
by Green-Kubo type formulas using the zero-momentum and zero frequency
limit of the imaginary part of the retarded equilibrium Greens
function of traceless components of the stress tensor.

We therefore need to understand first how the field theory stress
tensor can be obtained from the holographic dual. In holographic duals
the stress tensor is usually generated by a fluctuation of the metric
components. In the non-commutative theory it is important to
distinguish between the closed string metric, in terms of which the
supergravity solution is formulated and the open string metric, which
is the metric that is seen by open strings ending on a D-brane in a
B-field background \cite{Seiberg:1999vs}.

In this paper we will argue that the correct variables to look at in
order to define the gauge invariant operators in the dual field theory
are the open string variables: the open string metric, the open string
coupling and the $\Theta$-parameter that defines the star product on
the brane. The holographic stress tensor is therefore defined through
the fluctuations of the open string metric. As we will see it is
extremely important to realize this in order to be able to obtain a
well defined field theory stress tensor from the supergravity dual. In
fact it turns out that the temperature dependent part of the
holographic stress tensor calculated with the help of the open string
metric is precisely the same as the one in the usual commutative
$\Nfour$ theory.  Our findings are in agreement with the arguments
given in \cite{Li:1999am} and also in \cite{Arean:2005ar}

Since we define the stress tensor as the operator that is generated in
the holographic supergravity dual by open string metric fluctuations
it necessarily turns out to be symmetric.  A consequence of this is
that there are only two different Green-Kubo type formulas for the
viscosity coefficients
\begin{eqnarray}
  \eta_1 &=& -\Im \frac 1 \omega \int\, dt\, d^3x\, e^{i\omega t} 
\theta(t)\langle 
  [T_{yz}(t, \mathbf{x}) , T_{yz}(0,0) ]\rangle \,,\label{etauno}\\
  \eta_2 &=& -\Im \frac 1 \omega \int\, dt\, d^3x\, e^{i\omega t} 
\theta(t)\langle 
  [T_{xy}(t, \mathbf{x}) , T_{xy}(0,0) ]\rangle \,,\label{etados}
\end{eqnarray}
that are not related to each other by either symmetry of the stress
tensor or the $SO(2)$ symmetry that rotates $(y\leftrightarrow z)$.

In the following we will compute both of the above stress tensor
correlators and will also find the diffusion pole of the retarded
correlator of the momentum density operator in $y$ direction allowing
for gradients in the $x$ direction.  We advance here that all three
calculations end up in the universal value (\ref{eq:eta_universal}).

In the next section we present the supergravity background, establish
our conventions and define an effective five dimensional theory by
reduction on the $S^5$. In section three we compute the renormalised
temperature dependent part of the holographic stress tensor. Section
four is the core of this paper where we consider the various retarded
Greens function outlined above and show that the shear viscosity takes
the universal value (\ref{eq:eta_universal}) even in the anisotropic
situation presented by the non-commutative theory. We summarise our
conclusions in section five and outline possible interesting future
investigations.

\section{The background and dimensional reduction}

In \cite{Hashimoto:1999ut,Maldacena:1999mh} several geometries dual to
non-commutative N=4 supersymmetric Yang-Mills theory where proposed.
We shall concentrate on the one representing a finite temperature
quantum field theory with non-commutative plane along spacelike
directions $(y,z)$. In the string frame the metric is \be
ds^2_{10,string} = \HHH^{-1/2}\left( - f dt^2 + dx^2 + h(dy^2 +
  dz^2)\right) + \HHH^{1/2} (f^{-1} dr^2 + r^2 d\Omega_5^2 ) \,,
\label{stringmet}
\ee where \be f = 1- \frac{r_H^4}{r^4}~~~;~~~ h = \frac{1}{1 +
  \theta^2 \HHH^{-1}} ~~~;~~~ \HHH = \frac{L^4}{r^4} \,.  \ee The
AdS/CFT dictionary sets $L^4 = 4\pi \hat g N \alpha'^2$ where $\hat g
= g^2_{YM}$.  One easily sees that temperature and entropy density are
independent of the non-commutativity parameter $\theta$ \be T =
\frac{r_H}{\pi L^2}~~~~~~~;~~~~~~ s = \frac{N^2 \pi^2 T^3}{2}\,.  \ee
In order to ease comparison with previous calculations in the
literature we shall introduce new coordinates and definitions \be u =
\frac{r_H^2}{r^2}~~~;~~~~~ u_T =\frac{r_H^2}{L^2} = (\pi T
L)^2~~~~;~~~~ a = \theta u_T \, , \ee in terms of which the full
background acquires the following form, in the Einstein frame \beqa
ds^2_{10,E} &=&h^{-1/4} \left[\HHH^{-1/2}\left( - f dt^2 + dx_1^2 +
    h(dx_2^2 + dx_3^3)\right) + \frac{L^2}{f} \frac{du^2}{4 u^2} + L^2
  d\Omega_5^2 \right]\,,
\label{metd}
\\
f(u) &=& 1- u^2~~~;~~~ h(u) = \frac{u^2}{ u^2 + a^2} ~~~;~~~\HHH(u) =
\frac{u^2}{u_T^2} \,, \nonumber \eeqa and \beqa
e^{2\phi} &=&     h  \,,\nonumber\\
\rule{0mm}{6mm} H &=&\frac{a}{u_T} (h\HHH^{-1})' dy \wedge dz \wedge
dr\,,
\nonumber\\
\rule{0mm}{6mm} F_{(3)} &=& \frac{a}{u_T} (\HHH^{-1})' dt\wedge dx
\wedge dr\,,
\nonumber\\
\rule{0mm}{6mm} F_{(5)} &=& h (\HHH^{-1})' dt \wedge dx \wedge
dy\wedge dz \wedge dr + 4 L^4 \omega_{(5)} = ^*F_{(5)}\,.  \eeqa In
the rest of the paper we will work with an effective five dimensional
theory that is defined by dimensional reduction along the $S^5$
sphere.  Following \cite{Bremer:1998zp} we define a five dimensional
metric through
\begin{equation}
  ds_{10,E}^2 = e^{2\alpha \varphi} ds_{5}^2 + e^{2\beta \varphi} L^2 
d\Omega_5^2\,,
  \label{eq:5dmetric}
\end{equation}
where $\alpha = \sqrt{5/3}/4 $ and $\beta= - \sqrt{3/5}/4$ are chosen
such that $\varphi$ is a canonically normalised
scalar\footnote{\label{foot}On shell, the breathing mode $\varphi$ is
  related to the dilaton $\phi$.  Since $ds^2_{10,E} =
  e^{-\phi/2}ds^2_{10,string}$, from \eqref{stringmet}, \eqref{metd}
  and \eqref{eq:5dmetric} it follows that $e^{ 2\beta\varphi} =
  e^{-\phi/2} = h^{-1/4}$, or $\varphi = \sqrt{\frac{5}{3}}\phi $\,.}
and $ds_5^2$ is the five-dimensional Einstein frame metric.  The
action of this effective five dimensional theory is given
by\footnote{Our conventions are as follows $$ H = dB~~;~~~~F_{(1)} =
  dA_{(0)}~~;~~~~ F_{(3)} = dA_{(2)} + A_{(0)}\wedge H~~;~~~~ F_{(5)}
  = dA_{(4)} + \med A_{(2)}\wedge H - \med B\wedge d A_{(2)}\,.
 $$
}

\beqa S_{5} &=&= \frac{1}{2\kappa_5^2} \int d^5x
\left[\sqrt{-g_{5}}\left( {\cal R} - \frac{1}{2} (\partial \phi)^2 -
    \frac{1}{2} (\partial \varphi)^2 -
    \frac{1}{2} e^{2\phi} (\partial \chi)^2  - \frac{8}{L^2} e^{8 
\alpha\varphi} + e^{\frac{16}{5} \alpha \varphi} {\cal R}_5\right)  \right.  
\nonumber\\
&&\left. - \frac{1}{12} e^{\phi-4\alpha\varphi} F_{(3)}^2-
  \frac{1}{12} e^{-\phi-4\alpha\varphi} H^2 - \frac{2}{L} \left(
    A_{(2)} \wedge dB - d A_{(2)} \wedge B \right) \right] \,.
\label{eq:fiveDeffective}
\eeqa Here ${\cal R}_5$ denotes the curvature of the five-dimensional
sphere of radius $L$, i.e. ${\cal R}_5 = \frac{20}{L^2}$ and $
 \chi = - A_{(0)}$. The five-dimensional gravitational coupling is
given by $\kappa_5^2 = 4\pi^2 L^3/N^2$.  The resulting equations of
motion are given by \beqa
{\cal R}_{\mu\nu} &=& \frac{1}{2} \partial_\mu \phi\partial_\nu \phi+
\frac{1}{2} \partial_\mu \varphi\partial_\nu \varphi +\frac{1}{2} 
e^{2\phi}\partial_\mu \chi\partial_\nu \chi +   \frac{8}{3 L^{2}} 
e^{8\alpha \varphi} g_{\mu\nu} - \frac{1}{3}
e^{\frac{16\alpha}{5}\varphi} {\cal R}_{5} g_{\mu\nu} + \label{eqRmn} \\
& & \frac{1}{4} e^{\phi-4\alpha\varphi}\left((F_{(3)})^2_{\mu\nu} -
  \frac{2}{9} (F_{(3)})^2 g_{\mu\nu} \right) +
\frac{1}{4} e^{-\phi-4\alpha\varphi}\left( H^2_{\mu\nu} - 
\frac{2}{9} H^2 g_{\mu\nu} \right) \,,\\
\square \varphi &=& \frac{64\alpha}{L^2} e^{8\alpha\varphi} -
\frac{16}{5}\alpha e^{\frac{16\alpha}{5}\varphi} {\cal R}_{5} -
\frac{\alpha}{3} e^{\phi-4\alpha\varphi}(F_{(3)})^2 - \frac{\alpha}{3}
e^{-\phi-4\alpha\varphi} H^2
\,, \\
\square \phi &=& (\partial \chi)^2 e^{2\phi} + \frac{1}{12}
e^{\phi-4\alpha\varphi}(F_{(3)})^2 -
\frac{1}{12} e^{-\phi-4\alpha \varphi} H^2  \,,\\
d ( e^{2\phi} * d\chi) &=& -e^{\phi-4\alpha \varphi} H\wedge * F_{(3)} \,,\\
d ( e^{\phi-4\alpha\varphi}* F_{(3)}) &=& \frac{4}{L} H \,,\\
d ( e^{-\phi-4\alpha\varphi} * H ) &=& -e^{\phi-4\alpha\varphi
}d\chi\wedge * F_{(3)} - \frac{4}{L} F_{(3)} \,.
\label{eqvarphi}
\eeqa

\section{The holographic stress tensor}
The holographic stress tensor of the strongly coupled non-commutative
$N=4$ gauge theory has not been calculated until now in the
literature. In fact holography itself is poorly understood in this
background, mostly because the induced metric \eqref{eq:5dmetric} at a
fixed $r=\mathrm{const.}$ scales anisotropically as one goes with $r
\rightarrow \infty$ where it becomes degenerate. The authors of
\cite{Li:1999am} observed however that the anisotropy in the
dependence on $r$ encodes just the anisotropic decoupling limit
\cite{Hashimoto:1999ut, Maldacena:1999mh, Seiberg:1999vs} for the
metric components $g_{yy}$ and $g_{zz}$.  The same is true for the
dependence of the dilaton on $r$. If one therefore asks what are the
bulk fields that act as sources for the field theory operators it is
necessary to use the open string variables defined in
\cite{Seiberg:1999vs}
\begin{eqnarray}
  G_{\mu\nu} &=& g_{\mu\nu} - \left( B g^{-1} B \right)_{\mu\nu} \,, 
\label{eq:openG}\\
  \Theta^{\mu\nu} &=& 2 \pi \left( \frac{1}{g+B}\right)^{\mu\nu}_A \,,
\label{eq:Theta}\\
  G_s &=& g_s \left(\frac{\det G}{\det (g+B) } \right)^{\frac{1}{2}}\,,
\label{eq:openGs}
\end{eqnarray}
where the subscript $A$ denotes antisymmetrisation.  One can easily
check that, for the background given in \eqref{stringmet}, the open
string metric $G_{\mu\nu}$ is nothing but the usual (planar) AdS black
hole, the open string coupling $G_s$ is constant, and
$\Theta^{32}=\theta$.  In particular it follows that there is no
anisotropy in the open string metric.  Still, there is of course an
anisotropy in terms of the open string variables because of the non
vanishing $\Theta$ components which may lead to nontrivial physical effects (for example  on the drag force of a moving quark \cite{Matsuo:2006ws}).  We define the energy-momentum tensor
in the strongly coupled non-commutative field theory as the operator
that is sourced by the open string metric.  Therefore we have to find
out how the system responds to a variation of the open string metric
keeping the non-commutativity $\Theta$ and the open string coupling
$G_s$ fixed.

However the five dimensional action is formulated in terms of
$g^5_{\mu\nu}$, whereas the open string variables are written in terms
of the closed string metric.

The solution to $\delta \Theta =0$ gives the induced variations for
the B-field.  More explicitly:
\begin{equation}
  \delta \Theta^{\mu\nu} = -2\pi \left[\left(\frac{1}{g+B}\right) 
(\delta g + \delta B) \left(\frac{1}{g+B}\right) \right]^{\mu\nu}_A =0\,,
\end{equation}
and solving for $\delta B$ gives
\begin{equation}
  \delta B = \frac a u \left( \begin{array}{ccccc}
      0 & 0 &  -\delta g_{14} &  \delta g_{13} & 0 \\
      0 & 0 &  -\delta g_{24} &  \delta g_{23} & 0 \\
      \delta g_{14} &  \delta g_{24} & 0 & \frac{u^2  (\delta g_{33} + \delta 
        g_{44} )}{u^2-a^2} & \delta g_{45}\\
      - \delta g_{13} &  -\delta g_{23} & -\frac{u^2 (\delta g_{33} + \delta 
        g_{44} )}{u^2-a^2}  & 0 &  -\delta g_{35}\\
      0 & 0 &  -\delta g_{45} &  \delta g_{35} & 0 \\
    \end{array}  \right)\,.
  \label{eq.induced_variation_B}
\end{equation}
Similarly we find the induced variations on the dilaton. We set
$\delta G_s=0$, which is
\begin{equation}
  \delta G_s = G_s \left[ \delta \phi + \frac 1 2 (G^{-1})^{\mu \nu} 
\delta G_{\mu\nu} -
    \frac 1 2 \left(\frac{1}{g+B}\right)^{\mu \nu} (\delta g_{\mu\nu} + 
\delta B_{\mu\nu} )\right]=0\,,
\end{equation}
and find
\begin{equation}
  \delta \phi = -\frac{H^{1/2}}{2 h} \frac{a^2 }{(u^2-a^2)} 
( \delta g_{33} + \delta g_{44})\,.
\end{equation}
Notice that also the $S^5$ breathing mode $\varphi$ is linked, on
shell, with the dilaton (see footnote \ref{foot}). Therefore we also
have an induced variation of the form
\begin{equation}
  \delta \varphi = \sqrt{\frac{5}{3}}\,  \delta \phi\,.
\end{equation}

It is important to notice here that we have written the variations of
$B$ and $\phi$ in terms of the variations of the closed string metric
$g_{\mu\nu}=g^{10, string}_{\mu\nu}$. Now we need to relate these to
the variations of the five dimensional Einstein metric $g_5$. From
\eqref{eq:5dmetric} \be g^5_{\mu\nu} = e^{-2\alpha\varphi}
g^{10,E}_{\mu\nu} = e^{-2\alpha\varphi-\phi/2} g^{10,string}_{\mu\nu}
= e^{-4\phi/3}g_{\mu\nu} \,, \ee we obtain
\begin{equation}
  \delta g^5_{\mu\nu} = h^{-2/3}\left(\delta g_{\mu\nu}   - \frac 4 3  
g_{\mu\nu} \delta \phi\right)  \,.
\end{equation}

Following \cite{hstress} we define a quasilocal stress tensor in the
gravity theory as
\begin{equation}
  \tau^{\rho\lambda}_G = \frac{2}{\sqrt{-\Gamma}} 
\frac{\delta S}{\delta\Gamma_{\rho\lambda}}\,,
\end{equation}
where $\Gamma_{\mu\nu} = G_{\mu\nu} - \xi^r_{\mu} \xi^r_{\nu}$.  such
that $\xi^r_\mu$ is an outward looking normal vector to the
hypersurface $r=const$ fulfilling $\xi^r_\mu \xi^r_\nu G^{\mu\nu} =
1$.

Since we work with the effective five-dimensional theory
(\ref{eq:fiveDeffective}) we will first vary with respect to
$g^5_{\mu\nu}$. We therefore also need the restriction of the five
dimensional metric onto the hypersurface $r=const$. We will call this
$\gamma_{\mu\nu}$, so $\gamma_{\mu\nu} = g^5_{\mu\nu} -
\xi_{\mu}^{r,g}\xi_{\nu}^{r,g}$ and $\xi_{\mu}^{r,g}\xi_{\nu}^{r,g}
g^{5,\mu\nu} = 1$.

There are three contributions. The first is standard and comes from
the variation of the purely gravitational Einstein-Hilbert and
Gibbons-Hawking terms in the action
\begin{equation}
  \tilde\tau_g^{\rho\lambda} = \frac{1}{\kappa^2_5} \left( K^{\rho\lambda} - 
\gamma^{\rho\lambda} K^c_c \right)\,,
\end{equation}
where $K_{\mu\nu}$ is the extrinsic curvature of the metric that is
induced at $r=const$ by the fivedimensional metric $g^5_{\mu\nu}$.
The other contributions come from the induced variations on the
B-field and the dilaton and are given by
\begin{eqnarray}
  \tilde\tau_{B}^{\rho\lambda} &=& -\frac{1}{\kappa_5^2}\xi^r_\tau 
\left(e^{-\phi-4\alpha\varphi} H^{\tau\sigma\kappa} + \frac{1}{L} 
\epsilon^{\tau \mu\nu \sigma\kappa} A_{\mu\nu} \right)
  \frac{\delta B_{\sigma\kappa}}{\delta \gamma_{\rho\lambda}} \,,\\
  \tilde\tau_{\phi}^{\rho\lambda} &=& -\frac{1}{\kappa_5^2}\xi^r_\tau 
\partial^\tau \phi 
  \frac{\delta \phi}{\delta \gamma_{\rho\lambda}}  \,,\\
  \tilde\tau_{\varphi}^{\rho\lambda} &=& -\frac{1}{\kappa_5^2}
\xi^r_\tau \partial^\tau \varphi 
  \frac{\delta \varphi}{\delta \gamma_{\rho\lambda}} \,.
\end{eqnarray}
All the $\tilde\tau_{\mu\nu}$'s have been defined multiplied with a
factor of $\frac{2}{\sqrt{-\gamma^5}}$.  Now we need to convert
$\tilde\tau_g^{\mu\nu}$ into $\tau_G^{\mu\nu}$ according
to\footnote{We should actually compute $\delta \gamma^5_{\rho\lambda}
  / \delta \Gamma_{\mu\nu}$, but there is a projection implicitly in
  the definition of $\tau$ so that it is OK to use the full metrics
  here.}
\begin{equation}
  \tau_I^{\mu\nu} := \frac{\sqrt{-\gamma}}{\sqrt{-\Gamma}} 
\sum_{\lambda\leq\rho} \tilde\tau_I^{\rho\lambda} 
\frac{\delta g^5_{\rho \lambda}}{\delta G_{\mu\nu}}~~~~~;~~~~~
  I \in \{g,B,\phi,\varphi\}\,.
\end{equation}

The total stress tensor is now simply the sum of the different
contributions
\begin{equation}
  \tau_{\mathrm{total}}^{\mu\nu} =  
  ( \tau_G^{\mu\nu} +  \tau_B^{\mu\nu} + \tau_\phi^{\mu\nu} + 
\tau_\varphi^{\mu\nu} )\,. 
\end{equation}

Finally we define the vacuum expectation value of the stress tensor in
the field theory as
\begin{equation}
  \langle T^{ab} \rangle = \lim_{r\rightarrow\infty} \sqrt{-\Gamma} 
\xi^a_\rho \xi^b_\lambda \tau_{G,\mathrm{reg}}^{\rho\lambda} \,.
\end{equation}
In the last step a suitable regularization procedure has to be applied
on $\tau_{G,\mathrm{total}}$. In principle one would need to construct
local counterterms on the boundary to cancel the divergences as in
\cite{Balasubramanian:1999re}. In the case of the non-commutative
background these counterterms have not yet been constructed in the
literature. Therefore we employ the somewhat simpler method of
subtracting the gravity stress tensor of a reference spacetime before
taking the limit $r\rightarrow \infty$ \cite{hstress}. As reference
spacetime we take the zero temperature solution.  This is sufficient
for our purposes since we are only interested in the temperature
dependent part of the stress tensor.  We find 
\begin{equation}
  \langle T^{ab} \rangle  =  \frac{\pi^2 T^4 N^2}{8} 
\mathrm{diag}(3,1,1,1)\,,
\end{equation}
which is exactly the same result as for the commutative theory.  It is
worth emphasising the contrast between this result and the one found
in \cite{Cai:1999aw} where the stress-energy tensor of an
asymptotically flat background produced by a $D1-D3$ bound system was
computed and found to be anisotropic. In particular taking the naive decoupling limit in this solution leads to  negative pressures. It would be interesting to find an interpretation for this mismatch.

\section{Fluctuations}
The system of equations \eqref{eqRmn}-\eqref{eqvarphi} involves a
large number of fields. Therefore, looking for decoupled fluctuations
is now a more involved task, and calls for symmetry analysis.  In the
search for dispersion relations, that show up as poles in retarded
correlators, it is customary to assume a plane wave like perturbation
of the form $\Phi(u,k) = \phi(u) e^{i k x}$. For $k^\mu = (-\omega,
\vec k)$ this ansatz leaves a little group $O(2)$ of rotations in the
transverse plane. There is another $O(2)$ remnant of the isotropy
group that is broken by the background as soon as the
anticommutativity parameter $a\neq 0$ is switched on. When $\vec k$
has a component along the noncommutative plane the symmetry group is
completely broken. Perturbations involve typically all the fields and
become extremely tedious to handle.  The easiest situations occurs
when $\vec k = (k, 0 ,0)$, since in this case there is an overall
$O(2)$ symmetry group and the usual tensor decomposition applies. In
this paper, only such a family of plane waves will be considered.  We
will examine in detail, only two channels which directly measure the
shear viscosity: the scalar channel ($O(2)$ helicity $h=2$), and one
of the vector channels ($h=1$).  In this section $g_{\mu\nu}$ will
always stand for the five-dimensional Einstein frame metric.

\subsection{Scalar channel.}
When the polarization of the metric fluctuation lives inside the
non-commutative $(y,z)$-plane , $\delta g_{yz}$ rotates with $O(2)$ as
a tensor of spin 2 and decouples from the rest.  For such a
polarization we expect to find no poles on the retarded correlator.
Therefore, the usual alternative is to use Kubo formula, and for this,
just a time dependent fluctuations is enough $\delta g^y{_z} =
e^{-i\omega t} \rho_\omega(u)$. The equation that $\rho_\omega(u)$
satisfies is the usual one for a minimally coupled scalar \be
\rho_\omega'' - \frac{1+u^2}{u f} \rho_\omega' + \frac{\wn^2}{u f^2}
\rho_\omega = 0\,,
\label{minscal}
\ee whose solution, as an expansion in powers of $\wn =
\frac{\omega}{2\pi T}$ is the usual one \be \rho_\omega(u) =
(1-u^2)^{- \frac{i}{2}\wn}(1 + {\cal O}(\wn^2) +...)\,.  \ee The
normalisation $\rho_\omega(0)=1$ is in agreement with the fact that
the five-dimensional perturbation and the open-string perturbations
are related by $\delta g^y{_z} = \delta G^y{_z} $.  In a sense, this
anticipates the expected fact that in this channel no signal of the
non-commutative effects will show up.  It can be further checked by
computing the boundary gravitational action. A Fourier synthesis like
$\delta g^y{_z}(t,u) = \int \frac{d \omega}{2\pi} e^{-i\omega t}
f(\omega) \rho_\omega(u)$ may be plugged into the bare action \be S =
S_{5} + S_{GH} \label{bareact}\,, \ee where $S_{GH}=\frac{1}{8\pi
  G_5}\int_{u=\epsilon} d^4 x K$ stands for the usual Gibbons-Hawking
term.  After some algebra, the boundary action is found to be \be S=
\int d^3 x \frac{d\omega}{2\pi} \left.\rule{0mm}{5mm}
  f(\omega)f(-\omega){\cal F}(u)\right\vert^{u=1}_{u=\epsilon} \,, \ee
where the flux is given by \be {\cal F}(u) = -\frac{N^2}{8\pi^2
  L^3}\left(\frac{ u_T^2((3u^2-5)a^2+ u^2(u^2-3)}{L u^2(a^2+u^2)} -i
  \wn \frac{ u_T^2}{L} + ...\right)\,.
\label{fluxscal}
\ee The divergence at $u=0$ calls for boundary counterterms to be
added to the bare action \eqref{bareact}.  However it only shows up in
the real part! Using the holographic recipe for calculating retarded
Greens functions \cite{Son:2002sd} and the Green-Kubo formula
\eqref{etauno} we find \be \eta_1 = \lim_{\omega\to 0}
\frac{1}{\omega} \Im (-2 {\cal F}(u=0)) = \frac{N^2 \pi T^3}{8}\,, \ee
which leads to the universal value $\eta/s = 1/ 4\pi$ for the shear
viscosity to entropy ratio.

\subsection{Vector channel: dispersion relations}
Let us now choose $\delta g_x{^z}$. This perturbation has an index
along the propagation axis $x$ and another one, $z$, along the
non-commutative plane\footnote{This choice of index rising simplifies
  the equations, and affects the normalisation of the correlator, but
  does not modify the position of the searched poles.}. The
irreducible set of fluctuations involves six fields components, all
transforming as vectors of $O(2)$. Setting $\delta g_t{^z} =
e^{-i(\omega t + k x)}\rho_t, ~\delta g_x{^z} = e^{-i(\omega t + k
  x)}\rho_x$, $\delta A_{tz} = e^{-i(\omega t + k x)}u_T\alpha_t,
~\delta A_{xz} = e^{-i(\omega t + k x)}u_T\alpha_z$ and $\delta B_{ty}
= e^{-i(\omega t + k x)}u_T\beta_t, ~\delta B_{xy} = e^{-i(\omega t +
  k x)}u_T\beta_x$, the six fields
$\rho_x,\rho_t,\alpha_t,\alpha_z,\beta_t$ and $\beta_x$ satisfy a
system of six coupled second order ordinary differential equations,
plus three constraints. The system is not overdetermined and one can
find easily three linear differential relations that automatically
vanish. As explained in \cite{Kovtun:2005ev}, the natural variables to
look for, are combinations which are invariant under the residual
coordinate gauge transformations.  \beqa
Z &=& \qn\rho_t + \wn\rho_x \,,\\
V &=&\qn\alpha_t + \wn\alpha_x \,,\\
W &=& \qn\beta_t + \wn\beta_x \,.  \eeqa This set of variables
collapses into a system of 3 coupled differential equations.  Further
decoupling occurs if one defines the new set $(Z,V,W) \to (P,V,Q)$
with \be P = Z - \frac{a}{h} W~~~;~~~Q = W + \frac{a h}{u^2} Z\,.  \ee
Then we find a coupled system for $(V,Q)$ \beqa Q''-\frac{\qn^2
  f^2+(3u^2-1)\wn^2}{Ufa(\wn^2-f\qn^2)}Q'+\frac{u(\wn^2-f\qn^2)-4f}{u^2f^2}Q
-\frac{4\qn\wn}{f(\wn^2-f\qn^2)}V &=&0 \label{eqQ} \,,
\\
V''-\frac{\qn^2
  f^2+(3u^2-1)\wn^2}{Ufa(\wn^2-f\qn^2)}V'+\frac{u(\wn^2-f\qn^2)-4f}{u^2f^2}V
-\frac{4\qn\wn}{f(\wn^2-f\qn^2)}Q &=&0 \label{eqV}\,, \eeqa and a
decoupled equation for $P$ \beqa && P'' - h
\left(\frac{((u^2+1)\wn^2-f^2 \qn^2)}{Ufa(\wn^2- f \qn^2)} + a^2
  \frac{3\qn^2 f^2 + (5u^2-3)\wn^2)}{u^3 f(\wn^2- f \qn^2)}\right) P'+
\nonumber \\
&& h^2 \left(\frac{\wn^2- f\qn^2}{Ufa^2}+ a^2\frac{2u f^2 \qn^4 +
    4f(2u^4-4u^2-u\wn^2+2)\qn^2+2\wn^2(-2u^4+6u^2+\wn^2
    u-4)}{u^4f^2(\wn^2-f\qn^2)}\right.  \nonumber
\\
&& ~~~~~~~~~~~~~~~~~~~~~~~~~~~~~~~~~~~~~~~~~~~+\left.
  a^4\frac{f^2q^4-2f\wn^2\qn^2+\wn^2(\wn^2-4Ufa)}{u^5f^2(\wn^2-f\qn^2)}
\right)P =0 \,.  \eeqa The system \eqref{eqQ} and \eqref{eqV} is
independent of $a$. The natural combination to consider as a
gravitational perturbation is $P$, and we see that the equation for it
is explicitly dependent on $a$, signalling a possible influence of
this parameter in the final solution. However, as usual, we shall try
a perturbative solution in $\lambda$ \be P(u) = f(u)^{-i\wn/2}\left(
  P_0(u) + \lambda P_1(u) + ... \right)\,, \ee with $\wn\to
\lambda(-i\Gamma q^2), \qn \to\lambda\qn$, and normalised as
$P_0(1)=1, P_1(1)=0$.  One readily obtains \be P (u) =
f(u)^{-i\wn/2}\frac{1}{h(u)(a^2+1)}\left( 1 + i \frac{\qn^2}{2\wn}f(u)
  + {\cal O}(\wn^2, \qn^4,\wn\qn^2)\right)\,.  \ee It is helpful now
to remember the induced variations on the B-field in
(\ref{eq.induced_variation_B}).  Demanding as before $\delta \Theta =
0$ we find that in the new variables this is just $Q=0$.  Using this
and plugging the solution for $W$ back into the definition of $P$ we
find $P = Z/h$.  If we want to know the dispersion pole we need to set
$Z = h P |_{u=1}=0$. Notice that the overall normalisation of the
solution for $P$ or $Z$ is irrelevant here!  We observe therefore that
the hydrodynamic pole sits at the same position as in the commutative
case \cite{Kovtun:2005ev}.  \be \wn = -i\frac{\qn^2}{2}\,.  \ee This
is however not yet enough to infer that the shear viscosity is indeed
given by its universal value since the diffusion constant for momentum
diffusion is actually given by $\frac{\eta}{\epsilon + p}$. So far we
have only learnt that the diffusion constant is $D = \frac{1}{4\pi
  T}$.  Now it is important to know the energy momentum tensor in
equilibrium. As we have shown explicitly it is given by the same
expression as in the commutative case and obeys $\epsilon + p = T s$,
from which we can infer now the value of the shear viscosity as
\begin{equation}
  \frac{\eta_2}{s} = \frac{1}{4\pi}\,. \label{etados}
\end{equation}

 \subsection{Vector channel: Kubo formula}
 In the previous section we have computed the pole in the retarded
 correlation function of the gauge invariant variable $Z$. Since $Z$
 contains $\delta g^t{_z}$ and the latter is the source for the
 momentum density $\PP_z$ in $z$-direction what we really have
 computed is the diffusion constant for the process of momentum
 pointing along the non-commutative $z$-direction and diffusing into
 the commutative $x$-direction. By the remnant $SO(2)$ symmetry this
 is the same as the diffusion constant for the process of the
 $y$-momentum diffusing into $x$-direction.  On the other hand we have
 seen already, that viscous flow taking place in the non-commutative
 plane along is governed by the universality of the shear viscosity.
 This leaves us to determine the diffusion constant for the process of
 momentum density along the commutative direction  $p_y$ diffusing into a
 non-commutative direction $x$. To this end we will employ the Kubo
 formula \cite{McLennan}
 \begin{equation}
   \eta_3 = -\lim_{\omega\rightarrow 0} 
\frac{\Im G_{xy,yx}(\omega, 0)}{\omega}\,,
\label{dlkjl}
 \end{equation}
 where $ G_{xy,yx}=G_{xy,xy}$ is the retarded Greens function of
 $T_{xy}$. In fact the symmetry of the stress tensor defined through
 the variation of the open string metric implies immediately that
 $\eta_3=\eta_2$, so we should recover \eqref{etados}.  Nevertheless, it is an interesting exercise to
 recover the result of the previous section from the Kubo formula.
 Consider the purely time-dependent set of perturbations \be \delta
 g^x{_z} = \int \frac{d\omega}{2\pi} e^{-i\omega
   t}\Phi(\omega)\rho_\omega(u)~,~ \delta A_{tz}= u_T \int
 \frac{d\omega}{2\pi} e^{-i\omega t}\Phi(\omega)\alpha_\omega(u)~,~
 \delta B_{xy} = u_T \int \frac{d\omega}{2\pi} e^{-i\omega
   t}\Phi(\omega) \beta_\omega(u)\, .
 \label{pert}
 \ee One readily finds that $\alpha_\omega $ obeys a first order
 equation \be \alpha_\omega' =- \frac{2 a }{u^3}\rho_\omega -
 \frac{2}{u} \beta_\omega . \label{eqmoal} \ee Inserting this into the
 equations for $\rho$ and $\beta$, they can be casted as follows
 \begin{eqnarray}\label{eq:reducedsystem}
   \rho_\omega'' +\left( \frac{2h}{u}  -\frac{3-u^2}{u f}  \right) 
\rho_\omega' + \frac{\wn^2}{u f^2} \rho_\omega - \frac{2a h}{u} 
\left(\frac{2}{u f} \beta_\omega - \beta_\omega' \right) &=& 0 \, 
\label{eqmoone}\,,\\
   \beta_\omega'' + \left(\frac{2h}{u} - \frac{1+u^2}{u f} \right) 
\beta_\omega' + \left( \frac{\wn^2}{u f^2} - \frac{4h}{u^2 f} \right) 
\beta_\omega - \frac{2a h}{u^3} \rho_\omega' &=& 0\, .   \label{eqmotwo}
 \end{eqnarray}
 We expect four independent solutions to this system, two of which
 will be incoming at the horizon.  Parametrising the general solution
 by two constants, which we take to be the boundary values $\rho(0) =
 \rho_0$ and $\beta(0) = \beta_0$ we find, up to order ${\cal
   O}(\wn^2)$ \beqa \rho(u) &=& f(u)^{-i\wn/2}\frac{\rho_0 (a^2+ u^2)
   -\beta_0 u^2 a }{a^2 + u^2}\left( 1 - \frac{i\wn}{2} \frac{\rho_0
     u^2 a^2}{\rho_0 (a^2+ u^2) -\beta_0 u^2 a } \right)
 \label{solrho}\,,
 \\
 \beta(u) &=& f(u)^{-i\wn/2}\frac{\beta_0 a^2}{a^2 + u^2}\left( 1 -
   \frac{i\wn}{2} \frac{\rho_0 u^2}{\beta_0 a} \right)
 \label{solbeta}\,, \eeqa and from these, $\alpha(u)$ can be obtained
 integrating \eqref{eqmoal} \be \alpha(u)= f(u)^{-i\wn/2} \frac{\rho_0
   a f(u)}{u^2}\left( 1 + \frac{i\wn}{2} C_\alpha \frac{u^2}{\rho_0
     af(u)} \right)\,.  \ee Now, in order to proceed, we must compute
 the boundary action.  As usual, on shell, the regulated action $S=
 S_{Bulk}+ S_{GH}+ S_{ct}$ can be expressed in terms of boundary data
 \be S = \int d^3 x\frac{d\omega}{2\pi} ~\Phi(\vec x,\omega)
 \left(\rule{0mm}{5mm} \left.{\cal
       F}_{Bulk}(\omega,u)\right\vert^{u=1}_{u=\epsilon} + {\cal
     F}_{GH}(\omega, \epsilon) + {\cal F}_{ct}(\omega,
   \epsilon)\right)\Phi(\vec x,\omega) \,.  \ee We don't know the
 structure of counterterms. However, one can prove that the imaginary
 part of ${\cal F}_{Bulk}(\omega,u)+ {\cal F}_{GH}(\omega, u) $ is
 already a conserved flux, and hence independent of $u$.
 \begin{equation} \Im \left(\rule{0mm}{4mm}{\cal F}_{Bulk}(\omega,u)+
     {\cal F}_{GH}(\omega, u) \right) = -\wn\frac{
     u_T^2}{L}\frac{N^2}{8\pi^2 L^3}\left(
     \rho_0^2+(\beta_0^2+\rho_0^2)a^2-2\beta_0\rho_0 a\right)
   \label{fluxvec}\,.
 \end{equation}
 This is all we need in order to compute the shear viscosity, and
 conforms with the expectation that the hydrodynamic transport
 coefficients should not depend on the UV details of the theory.
 
 Comparing with \eqref{fluxscal}, we see that the resulting value of
 the shear viscosity depends on the correct choice of boundary
 conditions. Each one selects a corresponding quantum operator in the
 boundary theory.
 
 It is very interesting to compute the result corresponding to the
 operator $O_5$ that couples to the five dimensional closed string
 bulk metric $g_5$.  This can be done by simply setting
 $\rho_0=1,\beta_0=0$.  Plugging this into \eqref{fluxvec} and \eqref{dlkjl} we would end up
 with
 \begin{equation}
   \Im G_{O_5, O_5} = - \omega \frac{s }{4\pi} (1+a^2)\,.
 \end{equation}
 Using this would lead to a non-universal value for the shear
 viscosity  and moreover would be in conflict with the results in the
 previous section.

 Let us consider now the correct boundary stress tensor defined as the
 operator that couples to the open string metric. Using
 (\ref{eq.induced_variation_B}) it is easy to see that \be \beta =
 -\frac{a}{u^2}\rho\, .  \ee solves the constraint for varying only
 the open string metric.  For the solutions given in \eqref{solrho}
 and \eqref{solbeta}, the choice $\rho_0=0,\beta_0=1/a$ fulfils this
 condition. Using furthermore the definition of the open string metric
 (\ref{eq:openG}) it is not difficult to see that this corresponds
 precisely to an open string perturbation that obeys $\delta G^x{_z} =
 1$ at the boundary. Using this solution we find now for the
 imaginary part of the retarded correlator
 \begin{equation}
   \Im G(\omega, 0 ) =   - \omega \frac{s}{4\pi}\,.
 \end{equation}
 as expected.

 \section{Conclusions and Discussion}

 We have calculated the shear viscosity to entropy ratio in the
 holographic dual to a non-local non-commutative field theory.
 Although this theory is intrinsically anisotropic it turned out that
 the shear viscosities satisfy the conjectured bound exactly. We think
 this a rather remarkable result and collects further strong evidence
 in favour of the conjecture.  Although we were able to solve some
 important problems in this particular holographic theory there are
 many interesting questions that could be addressed in future work.

 One puzzling question is related to the membrane paradigm. It has
 been shown that hydrodynamic behaviour of black holes also arises in
 a purely gravitational perspective \cite{Kovtun:2003wp}. Explicit
 gravitational counterparts of shear modes have been constructed in
 the membrane paradigm approach.  In the non-commutative theory under
 consideration we emphasized the importance of the distinction between
 the open and closed string metric. From the pure gravity perspective
 however there is no reason a priori to consider the particular
 combination of fields defining the open string metric. From a bulk
 gravity perspective it is far more natural to consider the closed
 string metric.  Using therefore the membrane paradigm approach would
 suggest to consider closed string metric perturbations and it is far
 from clear if the hydrodynamic shear modes defined within the
 membrane paradigm using the bulk metric will give rise to the
 universal result (\ref{eq:eta_universal}). This points to a possible
 discrepancy between the membrane paradigm and holography!

 Another aspect concerns the holographic renormalization of the
 non-commutative theory. We have successfully implemented a simple
 subtraction scheme to compute a renormalised stress tensor. A much
 more powerful approach of holographic renormalization is to construct
 local covariant counterterms on the boundary. Such a program looks
 extremely difficult to realize if one considers the closed string
 variables, mostly because of the anisotropic behaviour in the
 holographic coordinate $u$ the non-commutative and commutative
 directions. We have pointed out that this anisotropy in the scaling
 with $r$ is somewhat fiducial since the correct variables to consider
 are the open string variables in which the metric is perfectly
 isotropic. This suggests that the holographic renormalization program
 of the non-commutative theory should be based on the the open string
 variables!

 A new difficulty arises when one switches on momentum $q$ in the
 non-commutative direction. It turns out that the differential
 equations for the linear fluctuations of the bulk fields are not any
 more of Fuchsian type but acquire essential singularities at the
 boundary. Moreover, many more fluctuations mix with each other than
 for the fluctuations considered in this paper. Therefore this
 presents as much a technical difficulty due to the high level of
 mixing as a conceptual one due to the presence of the essential
 singularities at the boundary.

 Finally we want to point out that the techniques developed in this
 paper should be sufficient to calculated the stress tensor and the
 stress tensor correlators at zero momentum also in other
 non-commutative backgrounds such as the non-commutative open string
 theory \cite{Gopakumar:2000na}.  We hope to make progress on these
 questions in future work.

 \acknowledgments The research of K.L. was supported in part by a
 Ram\'on y Cajal contract and by the Plan Nacional de Altas
 Energ\'{\i}as FPA-2006-05485 and by EC Commission under grant
 MRTN-CT-2004-005104.  The work of J.M. has been supported by MCyT,
 FEDER under grant FPA2005-00188, by EC Commission under grant
 MRTN-CT-2004-005104 and by Xunta de Galicia (Direcci\'on Xeral de
 Ordenaci\'on e Calidade do Sistema Universitario de Galicia, da
 Conseller\'\i a de Educaci\'on e Ordenaci\'on Universitaria).  We
 would like to thank P. Meessen, A. Starinets, A. Ramallo and M.A.
 V\'azquez-Mozo for useful discussion.

\end{document}